\documentclass{article}
\usepackage[T1]{fontenc}
\usepackage[utf8]{inputenc}
\usepackage[]{ismir} 
\usepackage{amsmath,cite,url}
\usepackage{graphicx}
\usepackage{color}
\usepackage{listings}
\usepackage{multirow}

\title{Exploring the Feasibility of LLMs for Automated Music Emotion Annotation}

\multauthor
  {Meng Yang$^1$ 
  \hspace{2mm}
  Jon McCormack$^1$ 
  \hspace{2mm}
  Maria Teresa Llano$^2$
  \hspace{2mm}
  Wanchao Su$^1$}
  {
  $^1$ SensiLab, Monash University, Melbourne, Australia\\
  $^2$ University of Sussex, Brighton, United Kingdom\\
  {\tt\small \{Meng.Yang, Jon.McCormack, Wanchao.Su\}@monash.edu, Teresa.Llano@sussex.ac.uk}
  }

\def\authorname{M. Yang, J. McCormack, T. Llano, and W. Su}

\begin{document}

\maketitle
\begin{abstract}
Current approaches to music emotion annotation remain heavily reliant on manual labelling, a process that imposes significant resource and labour burdens, severely limiting the scale of available annotated data. This study examines the feasibility and reliability of employing a large language model (GPT-4o) for music emotion annotation. In this study, we annotated GiantMIDI-Piano, a classical MIDI piano music dataset, in a four-quadrant valence-arousal framework using GPT-4o, and compared against annotations provided by three human experts. We conducted extensive evaluations to assess the performance and reliability of GPT-generated music emotion annotations, including standard accuracy, weighted accuracy that accounts for inter-expert agreement, inter-annotator agreement metrics, and distributional similarity of the generated labels. 

While GPT's annotation performance fell short of human experts in overall accuracy and exhibited less nuance in categorizing specific emotional states, inter-rater reliability metrics indicate that GPT's variability remains within the range of natural disagreement among experts. These findings underscore both the limitations and potential of GPT-based annotation: despite its current shortcomings relative to human performance, its cost-effectiveness and efficiency render it a promising scalable alternative for music emotion annotation.
\end{abstract}
\section{Introduction}\label{sec:introduction}


Music is widely recognized as a medium for conveying complex human emotions and experiences, making emotion-related research a focal point in the Music Information Retrieval (MIR) community \cite{MER}. Most empirical advances in emotion-related MIR start with a prerequisite step: securing a sufficiently large, reliable set of emotion labels. Existing datasets like DEAM \cite{DEAM}, Emotify \cite{Emotify}, VGMIDI \cite{VGMIDI} and EMOPIA \cite{Emopia}, were all created through intensive manual annotation campaigns. While indispensable, human labelling is slow and costly, so most of these datasets plateau at a few thousand items. These scale limits, in turn, the downstream research: modern deep architectures demand far more data than the community can currently afford to label by hand.
Recent advances in large language models (LLMs) have transformed text understanding, making it possible to infer music’s perceived emotion from extrinsic textual sources—metadata, lyrics, and contextual descriptions. Some of what shapes listeners’ perceived emotion is encoded outside the sound itself: composer biographies, genre conventions, and the historical context of composition \cite{portion}. Although LLMs cannot “hear” melody, harmony, or timbre, they can parse these documents and extract the affective stance they imply. In vocal music, lyrics already serve as an effective textual proxy and have underpinned successful MER studies \cite{lyrics1,lyrics2,lyrics3,lyrics4}. Instrumental works, however, lack built-in semantic cues; for them, metadata becomes the primary linguistic window into a composer’s expressive intent. This has been demonstrated in previous research, which has shown correlations between perceived emotion as well as historical and cultural context\cite{meta}, motivating our use of metadata‐driven LLM inference to annotate perceived emotion at scale for instrumental music.

In this study, we explore a novel annotation methodology that employs a large language model (GPT-4o) as an automated annotator for the perceived emotion of music. Our approach uses the title and composer of a music piece as search keywords to retrieve relevant web results, providing the LLM with extracted textual content as context, enabling it to infer an appropriate emotion label based on the available information. We apply our method to annotate GiantMIDI-Piano \cite{Giant_MIDI}, a classical piano dataset with 10,855 MIDI music pieces, using a four-quadrant valence-arousal framework. To evaluate this GPT-based annotation method, we randomly selected 100 samples from each of four emotion categories, totalling 400 samples, and obtained annotations for each sample from three human experts. We then compared the GPT-generated labels against the expert annotations using a comprehensive evaluation framework that includes binary and weighted accuracy metrics, inter-annotator reliability measures (Cohen’s Kappa and Fleiss’ Kappa), and distributional similarity analyses via Jensen–Shannon divergence.

Our findings show that although GPT-4o does not yet match human experts in overall accuracy or nuanced emotional categorization, its inter-rater variability falls within the range of natural disagreement among experts. These results highlight both the challenges and potential of LLM-based annotation: while further refinements are needed before it can fully replace manual annotation, the method’s cost-effectiveness and efficiency make it a promising approach for large-scale music emotion annotation.

The main contributions of this paper are as follows: (1) We propose a cost-effective approach to music emotion annotation by leveraging GPT’s text-based inference capability, reducing the reliance on time-consuming and costly manual labelling, (2) We develop a comprehensive evaluation framework incorporating accuracy metrics, inter-annotator agreement measures, and distributional analyses to assess and compare the performance of GPT-generated and expert annotations. 

\vspace{-5mm}

\section{Related Work}
\subsection{Music Emotion Annotation}

Early Music Information Retrieval (MIR) research on emotion, such as Music Emotion Recognition (MER), has relied heavily on manually annotated datasets. For example, CAL500 \cite{CAL500} contains 502 songs and each song is annotated with multiple human-provided emotion labels, and the DEAM dataset \cite{DEAM} includes 1,802 music excerpts with continuous and static arousal-valence annotations. While these corpora have proven invaluable for developing and evaluating models and tasks, manual emotion annotation requires multiple human listeners per track, making it both costly and labour-intensive \cite{labour,labour2,ACL}. As a result, most datasets are small in scale, typically comprising only hundreds or thousands of songs, and are often limited to specific genres \cite{CAL500,DEAM,Emotify,Emopia,VGMIDI}, 
constraining the performance on data-intense models. Models trained on such limited datasets struggle to generalize, and the lack of large-scale, standardized datasets complicates benchmarking across different studies.

Various alternative approaches have been proposed to address these challenges. Gamified annotation techniques such as MoodSwings \cite{MoodSwing}, and crowd-sourced tagging platforms like Last.fm and AllMusic, can expedite label collection, but they often introduce new problems such as data sparsity, biased sampling, or unclear taxonomy and a lack of label quality assurance \cite{labour}. 
The need for scalable, cost-effective, and consistent annotation methods has led researchers to explore advanced AI solutions -- including LLMs -- to assist or automate the labelling process.

\subsection{LLM-Based Annotation}
Recently, LLMs have revolutionized text-based annotation and classification tasks \cite{ACL_begin}. Unlike task-specific classifiers, LLMs are pre-trained on massive text corpora and can perform labelling through natural language prompts without task-specific retraining. Studies have shown that these models can match or even occasionally surpass the accuracy and consistency of crowd-sourced or expert annotations, primarily by applying labelling criteria more uniformly and reducing the impact of subjective interpretation \cite{crowd,CHI_crow}. Moreover, once deployed, LLMs annotate data rapidly and at relatively low cost, rendering them highly suitable for large-scale applications \cite{reduce_cost}.

Textual sources have long served as the source of emotional evidence for music–emotion studies. Early work inferred emotion directly from lyrics \cite{song_lyrics,lyrics1,lyrics2,lyrics3,lyrics4}, while more recent efforts have mined user-generated discourse—YouTube comments, tweets, Reddit threads—to annotate pieces along the valence–arousal plane, achieving moderate reliability with transformer models \cite{social}. These strategies, however, presuppose plentiful public discussion or vocal content and therefore miss much of the instrumental and lesser-known classical repertoire. Prior research shows that metadata—composer background, genre, stylistic school, and historical context—also correlates with music emotion \cite{meta}, which offers a broadly applicable foundation for automatic emotion annotation. Our study builds on this insight: we employ LLMs to annotate perceived emotion labels directly from contextual metadata, thereby expanding annotated resources for the non-lyrics music.

\vspace{-3mm}
\section{Methodology}\label{sec:typeset_text}

\subsection{Data Preparation}
We employed the GiantMIDI-piano dataset\cite{Giant_MIDI}, a classical piano MIDI collection comprising 10,855 files from 2,786 composers. For each piece, we collected contextual information by web-crawling a curated list of music information sources\footnote{e.g., \url{en.wikipedia.org}, \url{imslp.org}, \url{naxos.com}, \url{allmusic.com}, \url{classical-music.com}, \url{gramophone.co.uk}}.
We extracted metadata, such as genre, style, composer biography, historical and cultural context, and the composer's creative intent, which were then incorporated into the prompt provided to the GPT-4o for annotating the perceived emotion of the music.
\subsection{GPT-based Annotation}
\begin{figure}
  \centering
  \includegraphics[width=0.8\linewidth]{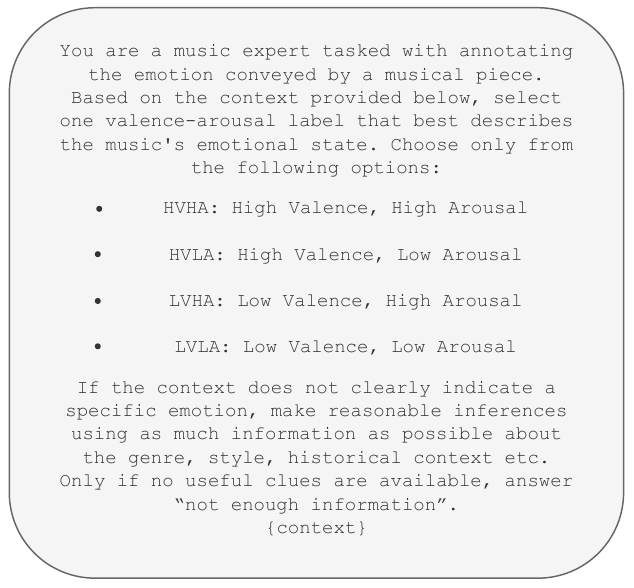}
  \vspace{-0.2cm}
  \caption{The Prompt for Music Emotion Annotation.}
  \label{fig:prompt}
  \vspace{-4mm}
\end{figure}

The collected text metadata was used to automate emotion annotation of the music in our dataset. Emotion labels were assigned according to Russell's valence–arousal model \cite{Russell}, following the quadrant-based scheme used in EMOPIA \cite{Emopia} to categorized into four discrete quadrants: High Valence–High Arousal (HVHA), High Valence–Low Arousal (HVLA), Low Valence–High Arousal (LVHA), and Low Valence–Low Arousal (LVLA). GPT-4o was supplied with contextual information via a structured prompt that explicitly instructed it to infer the perceived emotional content of each piece solely from the provided text, thereby minimizing hallucination. If the model determined that the available information was insufficient for a reliable annotation, it was instructed to return the label ``not enough information.'' The prompt is presented in Figure \ref{fig:prompt}. To ensure that GPT-4o selected the most reliable label based on the context and did not introduce random fluctuations, we set the model’s temperature to 0. Following the annotation process, a total of 9,803 musical pieces were assigned valid emotion labels.

\subsection{Human Evaluation}
To assess the feasibility and quality of GPT-generated annotations, we engaged three annotators with over five years of formal music training and similar cultural backgrounds. Before annotation, they calibrated their understanding of the valence–arousal framework to ensure a consistent labelling standard. We randomly sampled 100 tracks from each of the four emotion quadrants (400 total), ensuring a balanced distribution across quadrants. To reduce potential stylistic bias, the samples were selected to cover a diverse range of composers and musical styles within the classical piano repertoire. 
Each annotator independently labelled the perceived emotion based on listening experience. The annotations were then aggregated by majority voting to establish a human-derived \emph{gold standard}. Samples for which at least two out of three annotators agreed were designated as high-confidence samples, while those lacking consensus were classified as low-confidence samples. Our human annotation results yielded 386 high-confidence samples and 14 low-confidence samples.
\vspace{-2mm}
\subsection{Evaluation Framework}
We assessed the performance and reliability of the GPT-generated labels using a comprehensive evaluation framework that incorporates multiple metrics:
\vspace{-2mm}
\subsubsection{Accuracy}

Binary Accuracy is defined as the proportion of high-confidence samples for which the GPT-generated label exactly matches the gold standard obtained via majority voting among human experts. 

Given the subjectivity in music emotion annotation, a strict binary accuracy metric—where a sample is considered ``correct''  only if the GPT-generated label exactly matches the gold standard (i.e., the majority vote from experts)—may not capture the nuances of expert disagreement. To more precisely reflect the gradations in expert agreement, we propose a Weighted Accuracy that incorporates partial consensus among experts. Specifically, let $s_{i}$ be the score assigned to sample $i$ based on how closely GPT’s prediction aligns with expert consensus:

\begin{itemize}

    \item Full Consensus (3/3): All three experts agree on the same label. In this case if GPT's label matches the gold standard, $s_{i}=1$; Otherwise $s_{i}$ = 0.
    \item Partial Consensus (2/3): Two experts agree on a majority label $L_{m}$, and one expert has a minority label $L_{n}$. In this case if GPT's label equals $L_{m}$, $s_{i}$ = 1; if GPT's label equals $L_{n}$, $s_{i}$ = 0.5; Otherwise $s_{i}$ = 0\footnote{Note that each sample contributes at most one credit—so the weighted accuracy remains within $\left[ 0,1 \right]$, and GPT incurs no penalty when its annotation aligns with the majority.}.

    \item Complete Disagreement (3 distinct labels): All three experts disagree, each offering a unique label. If GPT's label matches any one of the three experts labels, $s_{i}$ = 1/3. Otherwise $s_{i}$ = 0.
 
\end{itemize}

Finally, the Weighted Accuracy is computed as the average score across all $N$ samples:
\vspace{-1mm}
\[
Weighted Accuracy = \frac{1}{N}\displaystyle\sum\limits_{i=1}^n s_{i}
\]

\vspace{-4mm}
\subsubsection{Inter-Annotator Consistency}
Cohen’s Kappa is used to measure pairwise agreement between two sets of annotations while accounting for chance agreement. For a pair of raters, it is given by:
\vspace{-2mm}
\[
\kappa = \frac{P_0 - P_e}{1 - P_e},
\]
where $P_0$ is the observed agreement and $P_e$ is the expected agreement by chance. We computed Cohen’s Kappa for GPT versus the gold standard, as well as for each pair of human annotators.

Fleiss’ Kappa extends the kappa statistic to multiple raters. Given a rating matrix $R$ where each row represents a sample and each column $j$ represents the number of ratings for category $j$, the per-item agreement is:
\vspace{-3mm}
\[
P_i = \frac{1}{n_i(n_i-1)}\displaystyle\sum\limits_{j=1}^k n_{ij}(n_{ij}-1)
\]
\vspace{-5mm}
with $n_i = \displaystyle\sum\limits_{j=1}^k n_{ij}$. The overall observed argreement is $\bar{P} = \frac{1}{N}\sum\limits_{i=1}^{N} P_i$, and the chance agreement is $P_e = \sum\limits_{j=1}^{k} p_j^2$, where $p_i$ is the proportion of ratings in category $j$ across all samples. Fleiss’ Kappa is then:
\vspace{-1mm}
\[
\kappa_F = \frac{\bar{P} - P_e}{1 - P_e},
\]
We computed Fleiss’ Kappa for both the human-only annotations and for the combined GPT and human annotations.

\subsubsection{Distributional Similarity}
JS divergence is used to measure the similarity between two probability distributions. 
For each sample, let $P=(p_1,...,p_k)$ be the aggregated expert distribution (derived from the relative frequencies of the three expert labels) and $Q = (q_1,...,q_k)$ be the one-hot representation of the GPT prediction. The JS divergence is defined as:
\vspace{-3mm}
\[
JS(P\parallel Q) = \frac{1}{2}KL(P\parallel M) + \frac{1}{2}KL(Q\parallel M)
\]
where $M=\frac{1}{2}(P+Q)$ and the Kullback–Leibler divergence $KL(P\parallel Q)$ is given by:
\vspace{-3mm}
\[
KL(P\parallel Q)= \displaystyle\sum\limits_{j=1}^k p_j\log\frac{p_j}{q_j}\;
\]
A lower JS divergence indicates greater similarity between the GPT-generated and aggregated expert distributions.

\vspace{-3mm}
\section{Experimental Results and Discussion}
\subsection{Pre-Test: Stability and Reproducibility of GPT-4o's Annotation}
Before conducting the main evaluation, we performed a pre-test to verify the stability and reproducibility of GPT-4o’s annotations under a deterministic setting (temperature = 0). This configuration ensures that the model consistently selects the most probable output, eliminating randomness and enabling precise evaluation. To test this, GPT-4o was prompted to annotate the same 400 samples across three independent runs. Results showed that 385 samples received identical labels in all runs, and the remaining 15 had two out of three consistent labels. This high level of consistency confirms that temperature = 0 yields stable, repeatable outputs, allowing us to attribute performance differences to the model’s underlying reasoning rather than stochastic variation.
\vspace{-2mm}
\subsection{Accuracy}
\subsubsection{Overall Annotation Performance}
\begin{table}[]
\small
\centering
\begin{tabular}{ccc}
\hline
Annotator & Accuracy & \begin{tabular}[c]{@{}c@{}}Weighted Accuracy\end{tabular} \\ \hline
GPT-4o                         & \textbf{0.710}                         & \textbf{0.788}                                                       \\ 
Human1                         & 0.833                         & -                                                            \\
Human2                         & 0.812                         & -                                                            \\
Human3                         & 0.869                         & -                                                            \\ \hline
\end{tabular}
\vspace{-3mm}
\caption{Accuracy Comparison of GPT-4o and Human Annotations with respect to the Gold Standard.}\label{acc1}
\vspace{-2mm}
\end{table}

\begin{table}[]
\small
\centering
\begin{tabular}{cc}
\hline
Model & Accuracy \\ \hline
GPT-3.5                    & 0.430    \\
GPT-4o                     & \textbf{0.710}    \\
GPT-4.5                    & 0.705    \\ \hline
\end{tabular}
\vspace{-2mm}
\caption{Accuracy Comparison among GPT models.}\label{gpt_model}
\vspace{-4mm}
\end{table}

\begin{table*}[!htb]
\centering
\small
\begin{tabular}{c|c|cccc}
\hline
Group             & N   & GPT-4o       & Human1      & Human2      & Human3      \\ \hline
Full-Consensus    & 211 & 180 (85.3\%) & 211         & 211         & 211         \\
Partial Consensus & 175 & 94 (53.7\%)  & 114 (65.1\%) & 115 (65.7\%) & 121 (69.1\%) \\ \hline
\end{tabular}
\vspace{-2mm}
\caption{Accuracy of GPT-4o and Human Annotators in Full and Partial Consensus Groups.}\label{acc_group}
\vspace{-3mm}
\end{table*}

Table \ref{acc1} demonstrates the accuracy of GPT-4o’s annotations against a human-derived gold standard, evaluated using both binary and weighted accuracy metrics. Binary accuracy, the proportion of high-confidence samples in which GPT-4o’s label exactly matched the gold standard, was approximately 71\%. In contrast, weighted accuracy, which accounts for varying degrees of expert consensus by awarding partial credit in cases of partial agreement, increased to around 78\%. Although human annotators achieved higher binary accuracy (ranging from approximately 81\% to 87\% across individual raters), the performance of GPT-4o is still acceptable considering the inherent subjectivity of music emotion annotation and the advantages in cost and efficiency.

\subsubsection{Comparison of GPT Model Performance}

We evaluated multiple GPT models from OpenAI for music emotion annotation. As results shown in table \ref{gpt_model}, the newly released GPT-4.5 achieves 70.5\% accuracy, matching GPT-4o’s 71\%, offering no notable improvement. In contrast, GPT-3.5 reaches only 43\%, underscoring significant advancements in the GPT-4 family's ability to accurately interpret and annotate musical emotion.

\subsubsection{Context Ablation Study}
To further validate the effectiveness of providing musical context, we conducted an ablation experiment in which GPT-4o was prompted to label the same 400 samples without any additional context information. In this “title-only” condition, GPT-4o had access to nothing more than the music title and composer name, and it achieved a binary accuracy of only 57\%. In contrast, when the model was furnished with the context we collected from online sources describing the work’s genre, stylistic background, historical and cultural factors, and composer biography, its accuracy rose significantly to 71\%. This gap underscores the importance of contextual information for disambiguating subtle emotional cues -- information that purely nominal references (e.g., a piece’s title) cannot reliably convey. Therefore, the contextual metadata can be seen as a crucial signal enabling GPT-4o to better align its annotations with expert judgments.

\subsubsection{Consensus-Based Subgroup Analysis}

In addition to overall accuracy metrics, we analyzed performance within two expert-consensus subgroups. As shown in table \ref{acc_group}, in the full-consensus group (211 samples in which all experts agreed), GPT-4o correctly annotated 180 samples (85.3\%) and erred on 31 samples. In the partial-consensus group (175 samples with a 2/3 expert agreement), GPT-4o matched the majority opinion in 94 samples (53.7\%), with no instances of aligning with the minority opinion and 21 samples classified entirely incorrectly. By comparison, human annotators in the partial-consensus subgroup achieved correctness ranging from 114 to 121 out of 175 samples (65.1\% to 69.1\%). These findings indicate that GPT-4o performs robustly when expert consensus is strong, and in cases of ambiguity, it captures a substantial portion of expert agreement.

\vspace{-2mm}
\subsection{Inter-Annotator Reliability Analysis}
Inter-annotator reliability was assessed using both Cohen’s Kappa and Fleiss’ Kappa. Table \ref{Pairwise} and table \ref{kappa_summary} present both pairwise Cohen’s Kappa and Fleiss’ Kappa values. The Cohen’s Kappa between GPT-4o and the gold standard (obtained via majority voting) was 0.613, indicating moderate agreement. Among the human experts, the pairwise Cohen’s Kappa values were 0.547, 0.568, and 0.569, with an average of 0.561. In addition, the pairwise Cohen’s Kappa values between GPT-4o and individual human annotators were 0.467, 0.593, and 0.607, yielding an average of 0.556. Although GPT-4o’s agreement with individual experts is slightly lower than the agreement observed among human annotators, the differences are relatively minor. At the group level, Fleiss’ Kappa was 0.561 for the human experts and 0.558 when GPT-4o was included, indicating that GPT-4o’s overall variability is comparable to that of the human raters.
\begin{table}[]
\small
\centering
\begin{tabular}{ccccc}
\hline
Annotator & GPT-4o & Human1 & Human2 & Human3 \\ \hline
GPT-4o    & N.A.   & 0.467  & 0.593  & 0.607  \\
Human1    & 0.467  & N.A.   & 0.547  & 0.568  \\
Human2    & 0.593  & 0.547  & N.A    & 0.569  \\
Human3    & 0.607  & 0.568  & 0.569  & N.A.   \\ \hline
\end{tabular}
\vspace{-2mm}
\caption{Pairwise Cohen's Kappa between GPT-4o and human experts.}\label{Pairwise}
\vspace{-3mm}
\end{table}
\begin{table}[]
\small
\centering
\begin{tabular}{ccccc}
\hline
\begin{tabular}[c]{@{}l@{}}GPT-4o \\ v.s. Gold\end{tabular} & \begin{tabular}[c]{@{}l@{}}Average \\ GPT-4o \\ v.s. \\ Human\end{tabular} & \begin{tabular}[c]{@{}l@{}}Average\\  Human \\ v.s. \\ Human\end{tabular} & \begin{tabular}[c]{@{}l@{}}Fleiss\\ Kappa \\ among \\ Human\end{tabular} & \begin{tabular}[c]{@{}l@{}}Fleiss \\ Kappa \\ GPT-4o v.s. \\ Human\end{tabular} \\ \hline
\multirow{2}{*}{0.613}                                   & \multirow{2}{*}{0.556}                                                  & \multirow{2}{*}{0.561}                                                    & \multirow{2}{*}{0.561}                                                   & \multirow{2}{*}{0.558}                                                       \\
                                                         &                                                                         &                                                                           &                                                                          &                                                                              \\ \hline
\end{tabular}
\vspace{-2mm}
\caption{Summary of inter-annotator reliability metrics in both average Cohen’s Kappa and Fleiss’ Kappa values.}\label{kappa_summary}
\vspace{-5mm}
\end{table}

\begin{table*}[]
\small
\centering
\begin{tabular}{ccc}
\hline
Comparison                                              & Main Difference & 95\% CI                      \\ \hline
GPT-4o vs. Gold minusHuman1 vs. Gold                       & -0.174          & {[}-0.250, -0.096{]}         \\
GPT-4o vs. Gold minus Human2 vs. Gold                      & -0.176          & {[}-0.241, -0.113{]}         \\
GPT-4o vs. Gold minus Human3 vs. Gold                      & -0.200          & {[}-0.267, -0.135{]}         \\ \hline
Fleiss' kappa, Human Experts minus GPT-4o \& Human Experts & 0.003           & \textbf{{[}-0.019, 0.025{]}} \\ \hline
\end{tabular}
\vspace{-2mm}
\caption{Bootstrap Differences in Pairwise and Group-Level Kappa.}\label{bootstrap}
\vspace{-2mm}
\end{table*}
\begin{table}[]
\small
\centering
\begin{tabular}{cc}
\hline
Comparison                  & Main Difference \\ \hline
GPT-4o v.s. Human1             & 0.266           \\
GPT-4o v.s. Human2             & 0.205           \\
GPT-4o v.s. Human3             & 0.194           \\ \hline
GPT-4o v.s. Aggregated Experts & \textbf{0.175}           \\ \hline
\end{tabular}
\vspace{-2mm}
\caption{Average Squared Jensen–Shannon Divergence between GPT-4o and Expert Annotation Distributions}\label{JS_divergence}
\vspace{-5mm}
\end{table}
To compare the agreement levels of GPT-4o against each individual expert, we performed bootstrap‐based hypothesis testing on the pairwise Cohen’s Kappa statistics. As shown in table \ref{bootstrap}, although the average Cohen kappa of GPT-4o versus gold (0.614) is close to the mean human kappa (0.593), bootstrap difference analysis reveals that GPT-4o's agreement with the gold standard is significantly lower than each expert, as evidenced by the 95\% confidence intervals for the differences (e.g.,[-0.267,-0.135] when comparing to Human3) all lying below zero. This outcome indicates that, on an individual basis, experts agree with the gold standard more consistently than GPT-4o does. However, these pairwise discrepancies should be interpreted in the context of normal inter‐rater variance: they highlight that GPT-4o tends to deviate from the gold standard more often than any single human annotator, rather than reflecting the overall group‐level consistency. Indeed, a subsequent Fleiss’ Kappa analysis reveals that incorporating GPT-4o into the set of raters does not significantly alter collective agreement—its mean difference from the human-only group is 0.003 with a 95\% confidence interval of [-0.019,0.025], indicating no significant impact. Consequently, while GPT-4o’s labeling differs more often from the gold standard than any single human expert, its variability at the group level remains within the range of human disagreement.

\subsubsection{Jensen–Shannon Divergence Analysis}
Table \ref{JS_divergence} shows the squared JS divergence for interpretability, where lower values indicate greater similarity between distributions. On an individual basis, the average squared JS divergence between GPT-4o and each expert was 0.266 (GPT-4o vs. Human1), 0.205 (GPT-4o vs. Human2), and 0.194 (GPT-4o vs. Human3). When comparing GPT-4o’s predictions to the aggregated expert distribution, the average squared JS divergence was lower at 0.175. These results suggest some divergence between GPT-4o’s predictions and individual expert labels, but the overall similarity to the combined expert consensus is moderate, indicating GPT-4o captures much of the collective expert opinion despite minor discrepancies with individual judgments.

\subsection{Error Analysis by Category}
\begin{figure}[t]
\centering
  \includegraphics[width=1\linewidth]{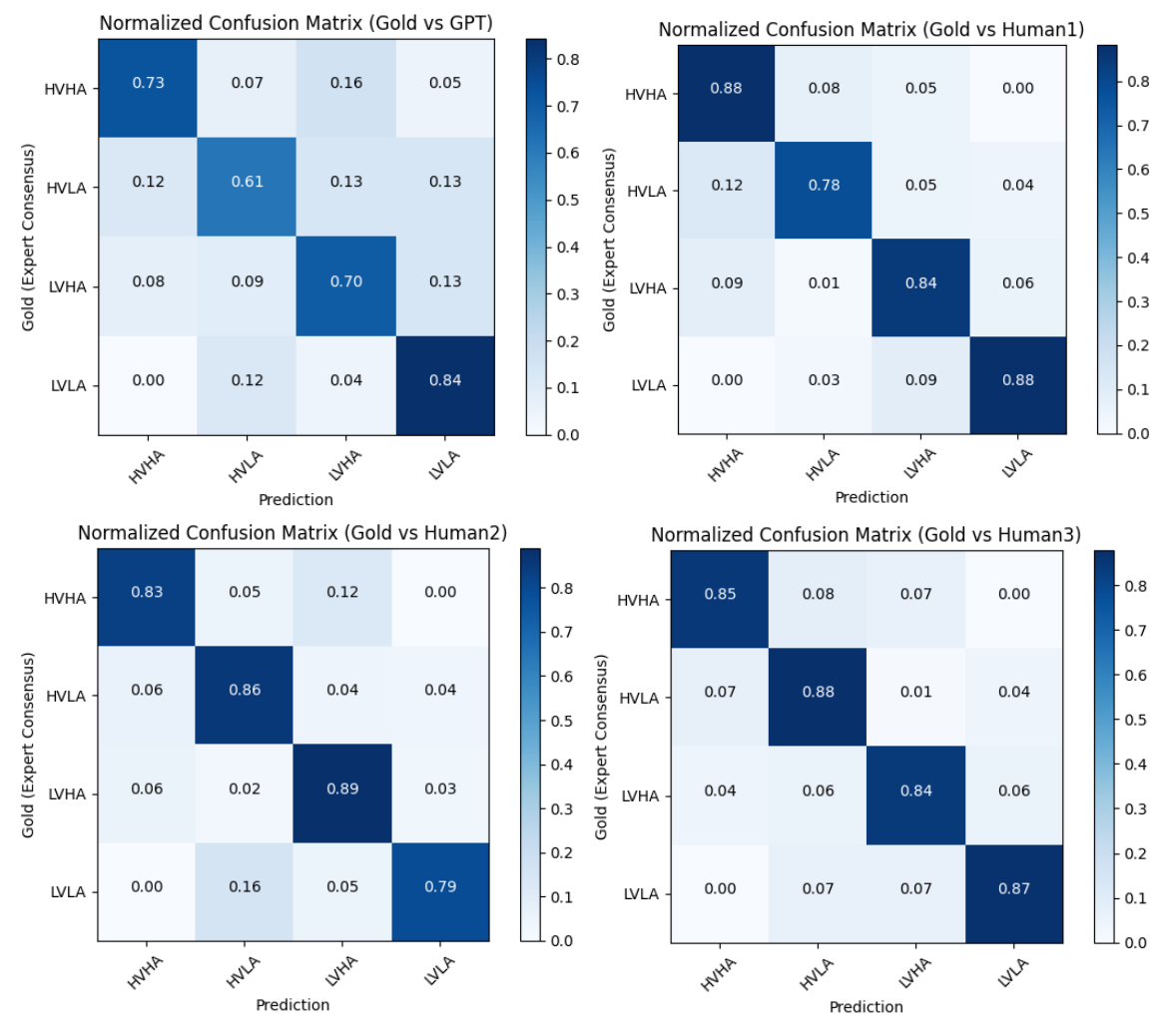}
  \vspace{-8mm}
  \caption{Normalized confusion matrix (by gold standard) comparing GPT-4o-generated emotion labels with expert consensus.}
  \vspace{-5mm}
  \label{confusion}
\end{figure}

Error analysis based on the normalized confusion matrices (Figure \ref{confusion}) reveals misclassification patterns for GPT-4o relative to the gold standard. In the high valence–high arousal (HVHA) category, for instance, GPT-4o correctly annotated 73\% of samples but misclassified about 16.0\% as low valence–high arousal (LVHA). A similar trend is observed in the low valence-low arousal (LVLA) category, where GPT-4o’s performance is relatively robust and even slightly exceeds that of Human2; however, GPT-4o tends to misclassify some LVLA samples as HVLA. These suggest that while GPT-4o is relatively adept at capturing the arousal dimension, it struggles to distinguish between positive and negative valence. Notably, This pattern of valence confusion is also present in Human2’s annotations, suggesting that even expert annotators may find it challenging to precisely distinguish the valence of some samples in this category. Additionally, GPT-4o exhibits suboptimal performance in the high valence–low arousal (HVLA) category, indicating a limited capacity to capture emotional nuances. Collectively, these findings highlight that while GPT-4o performs reasonably well in cases with clear emotional cues, its ability to discern fine-grained differences in emotion—particularly those involving the valence dimension—remains limited.

\vspace{-10mm}
\section{Discussion and Future work}
Our results indicate that while GPT-4o’s overall performance in music emotion annotation falls short of human expert-level accuracy, its variability remains within the range of natural disagreement among experts. Notably, GPT-4o tends to conflate emotional categories that differ primarily along the valence dimension, suggesting that valence is more challenging to infer from textual metadata than arousal, a difficulty also reflected in human annotations. This highlights a key limitation of our context-based method, in which GPT-4o annotations rely exclusively on pre-crawled textual metadata, therefore the accuracy and granularity of the annotations are limited by the quality and comprehensiveness of available textual information. Additionally, metadata-based inference rests on the majority cultural consensus encoded in text. While effective in most cases, this premise can mislabel “outlier” works whose affective intent departs from stylistic norms.


A critical issue arising from our findings relates to the subjectivity of music emotion annotation tasks. Music emotion can be approached from two perspectives: \textit{perceived} emotions (the emotions a listener believes the music is intended to convey) and \textit{induced} emotions (the emotions the listener personally experiences while listening)\cite{Gabrielsson}. Our methodology explicitly targets perceived emotions via textual contexts—cultural cues and genre conventions that shape listeners’ expectations before a note is heard—as a scalable proxy for annotation, reflecting a socially shared interpretation of what the music expresses rather than the idiosyncratic emotions it might induce. Even so, disagreement among human annotators reminds us that perceived emotion is not wholly objective, underscoring the inherent complexity in music emotion annotation.

Furthermore, relying on synthetic labels produced by large language models such as GPT-4o raises important ethical questions. Automated annotation undoubtedly offers scale and cost efficiency, yet risks eroding the nuanced interpretive judgments human experts provide. Replacing human annotation entirely could introduce systematic biases and oversimplifications—concerns that are especially salient in a domain as subjective as music–emotion research. We therefore advocate for a hybrid annotation framework to combine AI-generated initial annotations with human expert oversight for ambiguous cases. For instance, GPT-4o can serve as a first-pass annotator: high-confidence predictions—those backed by clear contextual cues—can be accepted directly after sampling validation, while low-margin or “not-enough-information” cases are routed to human annotators. This division of labour harnesses the scalability of LLMs while preserving essential human insight, and concurrently addressing broader ethical concerns about the responsible use of synthetic data.

Future work should focus on several key areas to address current limitations. First, refining prompt engineering by incorporating richer contextual details—such as comprehensive composer biographies, historical and cultural narratives, and program notes—may improve GPT-4o’s ability to capture emotional nuances, particularly along the valence dimension. Second, enabling real-time online context retrieval would allow the model access to more dynamic and detailed metadata, potentially improving its discriminative capabilities. Third, adopting the aforementioned hybrid human–AI annotation framework would balance efficiency and quality while also partially addressing the ethical concerns about AI-driven data annotation. Finally, exploring multi-modal approaches that integrate audio features with contextual metadata could provide a more holistic and accurate assessment of musical emotions, further bridging the gap between automated and expert-level annotations. Such models may also extend coverage to brand-new or undocumented works by enabling direct emotion inference from raw audio, complementing our current metadata-dependent pipeline.

Overall, although GPT-4o’s current performance does not entirely substitute for human expertise, its efficiency and scalability present a valuable opportunity to complement human annotations in large-scale Music Emotion research. Future work should refine prompting, enable online access, combine human–AI annotation, and integrate multimodal inputs to improve quality while addressing ethical and subjective challenges. 
\vspace{-4mm}
\section{Conclusion}
In this study, we presented a novel approach for automatic music emotion annotation using GPT-4o, leveraging contextual metadata extracted via web-crawling to label non-lyrical classical music. Our comprehensive evaluation, incorporating accuracy metrics, inter-annotator reliability measures, distributional similarity, and error analysis demonstrated that while GPT-4o does not yet match human expert accuracy, its outputs are stable and its variability falls within the range of natural human disagreement. Overall, our findings highlight the potential of LLM annotation as a scalable and cost-effective tool for large-scale music emotion annotation.
\bibliography{ISMIRtemplate}

\begin{thebibliography}{10}
\providecommand{\url}[1]{#1}
\csname url@samestyle\endcsname
\providecommand{\newblock}{\relax}
\providecommand{\bibinfo}[2]{#2}
\providecommand{\BIBentrySTDinterwordspacing}{\spaceskip=0pt\relax}
\providecommand{\BIBentryALTinterwordstretchfactor}{4}
\providecommand{\BIBentryALTinterwordspacing}{\spaceskip=\fontdimen2\font plus
\BIBentryALTinterwordstretchfactor\fontdimen3\font minus \fontdimen4\font\relax}
\providecommand{\BIBforeignlanguage}[2]{{%
\expandafter\ifx\csname l@#1\endcsname\relax
\typeout{** WARNING: IEEEtran.bst: No hyphenation pattern has been}%
\typeout{** loaded for the language `#1'. Using the pattern for}%
\typeout{** the default language instead.}%
\else
\language=\csname l@#1\endcsname
\fi
#2}}
\providecommand{\BIBdecl}{\relax}
\BIBdecl

\bibitem{MER}
Y.-H. Yang and H.~H. Chen, ``Machine recognition of music emotion: A review,'' \emph{ACM Transactions on Intelligent Systems and Technology}, vol.~3, no.~3, May 2012.

\bibitem{DEAM}
A.~Aljanaki, y.-h. Yang, and M.~Soleymani, ``Developing a benchmark for emotional analysis of music,'' \emph{PLOS ONE}, vol.~12, p. e0173392, 03 2017.

\bibitem{Emotify}
A.~Aljanaki, F.~Wiering, and R.~C. Veltkamp, ``Studying emotion induced by music through a crowdsourcing game,'' \emph{Information Processing \& Management}, vol.~52, no.~1, p. 115–128, Jan. 2016.

\bibitem{VGMIDI}
L.~N. Ferreira and J.~Whitehead, ``Learning to generate music with sentiment,'' in \emph{Proceedings of the Conference of the International Society for Music Information Retrieval}, Delft, Netherlands, 2019, pp. 384--390.

\bibitem{Emopia}
H.-T. Hung, J.~Ching, S.~Doh, N.~Kim, J.~Nam, and Y.-H. Yang, ``Emopia: A multi-modal pop piano dataset for emotion recognition and emotion-based music generation,'' in \emph{International Society for Music Information Retrieval Conference}, 2021.

\bibitem{portion}
M.~Barthet, G.~Fazekas, and M.~Sandler, ``Music emotion recognition: From content- to context-based models,'' in \emph{From Sounds to Music and Emotions}, M.~Aramaki, M.~Barthet, R.~Kronland-Martinet, and S.~Ystad, Eds.\hskip 1em plus 0.5em minus 0.4em\relax Berlin, Heidelberg: Springer Berlin Heidelberg, 2013, pp. 228--252.

\bibitem{lyrics1}
R.~Delbouys, R.~Hennequin, F.~Piccoli, J.~Royo-Letelier, and M.~Moussallam, ``Music mood detection based on audio and lyrics with deep neural net,'' \emph{ArXiv}, vol. abs/1809.07276, 2018.

\bibitem{lyrics2}
F.~H. Rachman, R.~Sarno, and C.~Fatichah, ``Music emotion detection using weighted of audio and lyric features,'' in \emph{2020 6th Information Technology International Seminar (ITIS)}, 2020, pp. 229--233.

\bibitem{lyrics3}
\BIBentryALTinterwordspacing
X.~Hu, K.~Choi, and J.~S. Downie, ``A framework for evaluating multimodal music mood classification,'' \emph{Journal of the Association for Information Science and Technology}, vol.~68, 2017. [Online]. Available: \url{https://api.semanticscholar.org/CorpusID:45480061}
\BIBentrySTDinterwordspacing

\bibitem{lyrics4}
Y.~Agrawal, R.~G.~R. Shanker, and V.~Alluri, ``Transformer-based approach towards music emotion recognition from lyrics,'' in \emph{Advances in Information Retrieval: 43rd European Conference on IR Research, ECIR 2021, Virtual Event, March 28 – April 1, 2021, Proceedings, Part II}, Berlin, Heidelberg, 2021, p. 167–175.

\bibitem{meta}
\BIBentryALTinterwordspacing
X.~Hu and J.~S. Downie, ``Exploring mood metadata: Relationships with genre, artist and usage metadata,'' in \emph{International Society for Music Information Retrieval Conference}, 2007. [Online]. Available: \url{https://api.semanticscholar.org/CorpusID:16794525}
\BIBentrySTDinterwordspacing

\bibitem{Giant_MIDI}
Q.~Kong, B.~Li, J.~Chen, and Y.~Wang, ``Giantmidi-piano: A large-scale midi dataset for classical piano music,'' \emph{Transactions of the International Society for Music Information Retrieval}, May 2022.

\bibitem{CAL500}
D.~Turnbull, L.~Barrington, D.~Torres, and G.~Lanckriet, ``Towards musical query-by-semantic-description using the cal500 data set,'' in \emph{Proceedings of the 30th Annual International ACM SIGIR Conference on Research and Development in Information Retrieval}, ser. SIGIR '07.\hskip 1em plus 0.5em minus 0.4em\relax New York, NY, USA: Association for Computing Machinery, 2007, p. 439–446.

\bibitem{labour}
\BIBentryALTinterwordspacing
P.~L. Louro, H.~Redinho, R.~Santos, R.~Malheiro, R.~Panda, and R.~P. Paiva, ``Merge -- a bimodal dataset for static music emotion recognition,'' 2025. [Online]. Available: \url{https://arxiv.org/abs/2407.06060}
\BIBentrySTDinterwordspacing

\bibitem{labour2}
Y.~E. Kim, E.~M. Schmidt, R.~Migneco, B.~G. Morton, P.~Richardson, J.~J. Scott, J.~A. Speck, and D.~Turnbull, ``Music emotion recognition: A state of the art review,'' in \emph{International Society for Music Information Retrieval Conference}, 2010.

\bibitem{ACL}
P.~Donnelly and A.~Beery, ``Evaluating large-language models for dimensional music emotion prediction from social media discourse,'' in \emph{Proceedings of the 5th International Conference on Natural Language and Speech Processing (ICNLSP 2022)}.\hskip 1em plus 0.5em minus 0.4em\relax Trento, Italy: Association for Computational Linguistics, Dec. 2022, pp. 242--250.

\bibitem{MoodSwing}
\BIBentryALTinterwordspacing
Y.~E. Kim, E.~M. Schmidt, and L.~Emelle, ``Moodswings: A collaborative game for music mood label collection,'' in \emph{International Society for Music Information Retrieval Conference}, 2008. [Online]. Available: \url{https://api.semanticscholar.org/CorpusID:14382686}
\BIBentrySTDinterwordspacing

\bibitem{ACL_begin}
Z.~Tan, D.~Li, S.~Wang, A.~Beigi, B.~Jiang, A.~Bhattacharjee, M.~Karami, J.~Li, L.~Cheng, and H.~Liu, ``Large language models for data annotation and synthesis: A survey,'' in \emph{Proceedings of the 2024 Conference on Empirical Methods in Natural Language Processing}.\hskip 1em plus 0.5em minus 0.4em\relax Miami, Florida, USA: Association for Computational Linguistics, Nov. 2024, pp. 930--957.

\bibitem{crowd}
F.~Gilardi, M.~Alizadeh, and M.~Kubli, ``Chatgpt outperforms crowd workers for text-annotation tasks,'' \emph{Proceedings of the National Academy of Sciences of the United States of America}, vol. 120, 2023.

\bibitem{CHI_crow}
Z.~He, C.-Y. Huang, C.-K.~C. Ding, S.~Rohatgi, and T.-H.~K. Huang, ``If in a crowdsourced data annotation pipeline, a gpt-4,'' in \emph{Proceedings of the 2024 CHI Conference on Human Factors in Computing Systems}, ser. CHI '24.\hskip 1em plus 0.5em minus 0.4em\relax New York, NY, USA: Association for Computing Machinery, 2024.

\bibitem{reduce_cost}
S.~Wang, Y.~Liu, Y.~Xu, C.~Zhu, and M.~Zeng, ``Want to reduce labeling cost? {GPT}-3 can help,'' in \emph{Findings of the Association for Computational Linguistics: EMNLP 2021}.\hskip 1em plus 0.5em minus 0.4em\relax Punta Cana, Dominican Republic: Association for Computational Linguistics, Nov. 2021, pp. 4195--4205.

\bibitem{song_lyrics}
D.~Edmonds and J.~Sedoc, ``Multi-emotion classification for song lyrics,'' in \emph{Proceedings of the Eleventh Workshop on Computational Approaches to Subjectivity, Sentiment and Social Media Analysis}.\hskip 1em plus 0.5em minus 0.4em\relax Online: Association for Computational Linguistics, Apr. 2021, pp. 221--235.

\bibitem{social}
\BIBentryALTinterwordspacing
P.~Donnelly and A.~Beery, ``Evaluating large-language models for dimensional music emotion prediction from social media discourse,'' in \emph{Proceedings of the 5th International Conference on Natural Language and Speech Processing (ICNLSP 2022)}, M.~Abbas and A.~A. Freihat, Eds.\hskip 1em plus 0.5em minus 0.4em\relax Trento, Italy: Association for Computational Linguistics, dec 2022, pp. 242--250. [Online]. Available: \url{https://aclanthology.org/2022.icnlsp-1.28/}
\BIBentrySTDinterwordspacing

\bibitem{Russell}
J.~Russell, ``{A circumplex model of affect},'' \emph{Journal of personality and social psychology}, vol.~39, no.~6, pp. 1161--1178, 1980.

\bibitem{Gabrielsson}
A.~Gabrielsson, ``Emotion perceived and emotion felt: Same or different?'' \emph{Musicae Scientiae}, vol.~5, no. 1\_suppl, pp. 123--147, 2001.

\end{thebibliography}

%
%
%
%

\end{document}